\documentclass[12pt]{article}
\usepackage{amsmath}
\usepackage{amssymb}
\usepackage{graphics}
\newcommand{\tr}{\operatorname{tr}}
\newcommand{\opname}[1]{\operatorname{#1}}
\newcommand{\Slash}[1]{{#1}\negthickspace{\negthickspace{\slash}}}

\begin{document}

\title{Quantum Mechanics of Lowest Landau Level\\ 
Derived from $\mathcal{N}=4$ SYM with Chemical Potential}
\author{D.~Yamada\footnote{dyamada@u.washington.edu}\\
\textit{Department of Physics, University of Washington,
Seattle, WA. 98195}}
\date{}
\maketitle

\begin{abstract}
The low energy effective
theory of $\mathcal{N}=4$ super-Yang-Mills theory on $S^3$ with
an $R$-symmetry chemical potential is shown to be the lowest Landau
level system. 
This theory is a holomorphic complex matrix
quantum mechanics.
When the value of the chemical potential is not far below the mass
of the scalars, the states of the effective theory consist only
of the half-BPS states.
The theory is solved by the operator
method and by utilizing the lowest Landau level projection
prescription for the value of the chemical potential less than
or equal to the mass of the scalars. When the chemical potential
is below the mass, we find
that the degeneracy of the lowest Landau level is lifted and 
the energies of the states are computed. The one-loop correction
to the effective potential is computed for the commuting fields
and treated as a perturbation
to the tree level quantum mechanics.
We find that the perturbation term has non-vanishing matrix elements that
mix the states with the same $R$-charge.
\end{abstract}

\pagebreak
\tableofcontents


\section{Introduction}
The half-BPS sector of the AdS/CFT correspondence 
\cite{Maldacena:1997re,Gubser:1998bc,Witten:1998qj}
has recently been 
actively investigated since the works of Berenstein \cite{Berenstein:2004kk}
and Lin, Lunin and Maldacena \cite{Lin:2004nb} (LLM). The former noted
that a certain decoupling limit of $\mathcal{N}=4$ super-Yang-Mills
theory (SYM) on $\mathbb{R}\times S^3$ singles out the half-BPS
states of the theory and the theory is simply 
described by the one dimensional fermionic harmonic oscillators. 
This motivated
the latter authors to construct the geometry of Type IIB supergravity 
corresponding to the half-BPS states of the field theory. 
Remarkably, they were able to 
obtain the generic solution that describes all the half-BPS states.
The geometries are determined by solving a function that satisfies
a partial differential equation. This procedure requires to specify
a boundary condition on a particular two dimensional plane in
ten dimensions and LLM conjectured that the boundary condition
on the plane is identified to the phase space of the corresponding
fermion system in the field theory.

On the field theory side, a half-BPS operator is characterized
by its dimension being equal to its $R$-charge. Under the
conformal mapping of the theory onto a sphere, the dimension
is mapped to the energy of the corresponding state measured
in the units of the inverse radius of the sphere. 
When we introduce a chemical potential corresponding to the
$R$-symmetry, the Hamiltonian of the system takes the form,
\begin{equation*}
	H=h-\mu j
			\;,
\end{equation*}
where $h$ is the original Hamiltonian, $\mu$ is the chemical
potential and $j$ is the $R$-charge number operator.
When the chemical potential is tuned to exactly $1$, which
is measured in the units of the inverse radius $R$, any half-BPS
states, $|\psi\rangle$, satisfy the condition,
\begin{equation*}
	H|\psi\rangle=(h-j)|\psi\rangle=0
						\;.
\end{equation*}
It was pointed out that such behavior of the half-BPS sector
is analogous to the Landau problem.
To the knowledge of the author, this
was first noted by Berenstein \cite{Berenstein:2004kk}.

In the present paper, we consider $\mathcal{N}=4$ 
SYM on a sphere with
the chemical potential, $\mu$, and show that the states of
the low energy effective theory with $\mu\sim 1$ are the 
half-BPS states. This gives a clear physical picture of the
decoupling limit that isolates the half-BPS states from
other states of the theory. The Hamiltonian of the low
energy effective theory is the one that describes the
lowest Landau level system and it is a holomorphic
complex matrix quantum mechanics.
The attempt
to make the precise correspondence between the SYM and the Landau 
problem has been made by Ghodsi \textit{et al.} in 
\cite{Ghodsi:2005ks}. However, we would like to emphasize
that we are going to directly derive the lowest Landau level
system as a low energy effective theory of the SYM with the
chemical potential rather than developing the dictionary 
between the two.

One may start by asking a few questions. 
First, what happened to the oppositely charged particles?
In a relativistic field theory, a particle is necessarily accompanied
by the corresponding anti-particle and it has the opposite charge. Therefore,
the particle and anti-particle respond differently to the chemical potential
applied to the system. In the case of Minkowski space, the dispersion
relations appear as $E\pm\mu$,
so the above statement, $(h-j)|\psi\rangle=0$,
seems to correspond to either a particle
or an anti-particle (in below, we take it as a particle as our convention)
and not the both together. 
They differ in energy by the amount $2$ when the chemical potential is $1$.
This suggests that one should be considering
the low energy effective theory of $\mathcal{N}=4$ SYM on sphere
with the scale $1/R$. In such
low energy effective theory, the anti-particle excitations are too
costly and they are always at the ground state. In the language of
the Landau problem mentioned above, one should consider the system
projected down to the lowest Landau level. As we shall see, 
considerations along this
line lead to the clarification of the decoupling limit that
isolates the half-BPS states and reveal the emergence of the
holomorphic complex matrix quantum mechanics.

The second question that one may ask is the response of the system
when the chemical potential is slightly off from the value $1$.
When the chemical potential satisfies $\mu>1$, it effectively introduces
negative mass terms in the scalar fields and the precise determination
of the outcome requires higher loop computations. The system may
simply become unstable and cease to exist or it may go through
a phase transition. In the present work, we limit our analysis to
the tree level and to the one-loop with the special case where the fields
are assumed to take values in flat directions
and do not deal with this
question. (However, see \cite{YY:2005} for the high temperature
analysis of such case where the Higgs phase transition has been found.)

When we have the chemical potential slightly smaller than the value
$1$, we expect the degeneracy of the lowest Landau level to be lifted,
and the half-BPS states respond to the chemical potential by
$H|\psi_n\rangle=(1-\mu)n|\psi_n\rangle$, where $n$ is the charge
of the state. In the free field theory,
we do not expect the stress of the chemical potential to mix
the half-BPS states and further, the charge conservation
requires the states with different charges not to mix.
When we turn on the interaction of the theory, we should expect
that the states with the same $R$-charge mix to form an energy
eigenstate.

In what follows, we deal with the questions addressed above.
In Section \ref{effthy}, we construct the tree level effective
action of $\mathcal{N}=4$ SYM on $\mathbb{R}\times S^3$ with
chemical potentials corresponding to the $R$-symmetry. In addition,
we present the one-loop correction to the effective action for the
special case where the fields of the effective theory are taking
values in the flat directions. The details of the one-loop
computation is given in Appendix \ref{AppxOneLoop}.
In
Section \ref{QM}, we solve the non-interacting system first 
by the operator method.
Then, we rearrange the Hamiltonian of the system resulting from
the effective action and show
explicitly that the Hamiltonian is exactly that of the Landau
problem plus a potential that vanishes when $\mu=1$.
We solve the system again by utilizing the lowest 
Landau level projection
prescription of Girvin and Jach \cite{Girvin:1984}. 
Here, we find
that the half-BPS states respond to the chemical potential as
expected above and we show that the tree level interaction does 
not mix the states,
while the one-loop correction does mix the states with the same
$R$-charges.
In Section \ref{Interpretation}, we discuss
the result in the light of the AdS/CFT correspondence mentioned
at the beginning of this section,
in particular, the non-commutativity of coordinates
is discussed from the
perspective of the lowest Landau level projection prescription.
We conclude by discussing the further directions
in Section \ref{Outlook}.


\section{The Effective Action with Chemical Potential}\label{effthy}
We consider $U(N)$ $\mathcal{N}=4$ SYM on $\mathbb{R}\times S^3$ where
the number of the color $N$ is not assumed to be large.
Our aim in this section is to obtain the low energy effective
action of the theory with chemical potentials where the scale of the
effective theory is $1/R$ with $R$ being the radius of
the sphere. This theory has
scalar mass terms that arise from the conformal scalar-curvature
coupling and for the case of $S^3$, the mass of the scalars 
are the inverse of the radius $R$. The presence of
these mass terms motivates us to
introduce chemical potentials to the theory where, as we will see,
chemical potentials appear as negative mass-squared terms in the
Lagrangian.

It is known and also described in \cite{YY:2005} that given a non-Abelian
global symmetry of a theory, one can introduce chemical potentials 
corresponding to a Cartan subalgebra of the symmetry. In 
our case, we have the global $SU(4)_R$ $R$-symmetry, so
we can introduce three chemical potentials to the theory corresponding
to $U(1)^3$ subgroup of the $R$-symmetry.

In order to write down the Lagrangian, we first would like to determine
how the chemical potentials are assigned for each field of the
theory. The field content of the theory is six scalars, four (left-handed)
Weyl fermions and a vector,
all in adjoint representation of the
gauge group $U(N)$. The table below summarizes the transformation laws
of the fields under the $SU(4)_R$.
\\
\\
\centerline{
\begin{tabular}{c|c c}
Fields & DOF & $SU(4)_R$ \\ \hline
$\phi$ & $6N^2$       & $\mathbf{6}$	\\ 
$\lambda$ & $8N^2$        & $\mathbf{4}$ \\
$A$    & $2N^2$          & $\mathbf{1}$
\end{tabular}}
\\
\\
Here we have equal number of bosonic and fermionic degrees of freedom
as usual in supersymmetric theories. The six scalars $\phi$ are actually
complex, but with three complex relations, leaving six real degrees
of freedom. To see this, consider the set of six complex scalars, $\phi_{ij}$
with $i,j=1\ldots,4$ and $\phi_{ij}=-\phi_{ji}$,
transforming under the anti-symmetric representation
$\mathbf{6}$ of $SU(4)_R$. The complex conjugate
representation $\bar{\mathbf{6}}$ is given by,
\begin{equation*}
	\phi^{ij}=\frac{1}{2}\epsilon^{ijkl}\phi_{kl}
								\;,
\end{equation*}
and we must impose $\phi^{ij}=\phi_{ij}^*$. These constraints lead to
the relations,
\begin{equation}\label{ComplexScalarRelations}
	\phi_{12}^*=\phi_{34}
							\;,\quad
	\phi_{13}^*=\phi_{42}
							\;,\quad
	\phi_{14}^*=\phi_{23}
							\;,
\end{equation}
thus, as claimed, leaving six real degrees of freedom in the
scalar sector.

We choose a Cartan subalgebra of the defining 
representation $\mathbf{4}$ as,
\begin{align*}
	Q^\mathbf{4}_1=\frac{1}{2}\opname{diag}(1,1,&-1,-1)\;,\quad
	Q^\mathbf{4}_2=\frac{1}{2}\opname{diag}(1,-1,1,-1)\;,\\
	Q^\mathbf{4}_3&=\frac{1}{2}\opname{diag}(1,-1,-1,1)\;.
\end{align*}
One can obtain the anti-symmetric representation $\mathbf{6}$, by considering the
transformations of an anti-symmetric 2-tensor under $\mathbf{4}\otimes\mathbf{4}$.
We choose a basis so that the matrices of $\mathbf{6}$ appear as,
\begin{align*}
	Q^\mathbf{6}_1=\opname{diag}(1, -1,&0, 0, 0,0)\;,\quad
	Q^\mathbf{6}_2=\opname{diag}(0, 0, 1, -1, 0, 0)\;,\\
	Q^\mathbf{6}_3&=\opname{diag}(0, 0, 0,0, 1, -1)\;.
\end{align*}
Then, the introduction of the chemical potentials modifies the Lagrangian
by the replacement of the covariant derivatives,
\begin{equation}\label{CovariantDerivative}
	D_\nu\rightarrow D_\nu-\mu_jQ_j\delta_{\nu,0} 
							\;,
\end{equation}
where $\mu_j$ are the chemical potentials and 
we assumed the Euclidean signature here.
One notices that the time component of the covariant derivative is modified 
by the chemical potentials.
One can see this by the spurion argument.
Imagine gauging the $U(1)$ global symmetry
that you introduce a chemical potential. The chemical potential couples to the
conserved $U(1)$ charge, just as the zeroth component of the minimal coupling.
So it can be thought as a background gauge field whose components
are all zero except the time component. Then, this gauged vector boson must
appear in the time component of the covariant derivative as in above.

The modified covariant derivative (\ref{CovariantDerivative}) shows
that the chemical potentials are assigned for the four Weyl spinors as,
\begin{align*}
	&\tilde{\mu}_1:=\frac{1}{2}(\mu_1+\mu_2+\mu_3),\;
	\tilde{\mu}_2:=\frac{1}{2}(\mu_1-\mu_2-\mu_3),\; \nonumber\\
	&\tilde{\mu}_3:=\frac{1}{2}(-\mu_1+\mu_2-\mu_3),\;
	\tilde{\mu}_4:=\frac{1}{2}(-\mu_1-\mu_2+\mu_3).\;
\end{align*}
and the chemical potentials, $\pm\mu_p$ with $p=1,\ldots,3$
are given to the scalars $\phi_{1,p+1}$ 
and their conjugates as in (\ref{ComplexScalarRelations}).

We re-express the Weyl fermions in terms of Majorana fermions.
In four dimensions,
the kinetic terms of the Weyl fermions, $\lambda_i$,
with chemical potentials are related
to those of Majorana fermions, $\psi_i$, by,
\begin{equation*}
	\lambda_i\tau_\nu 
	(\stackrel{\leftrightarrow}{D}_\nu-\tilde{\mu}_i\delta_{\nu,0})\bar{\lambda}^i
	=\bar{\psi}_i(\Slash{D}+\tilde{M}_i)\psi_i	\;,
\end{equation*}
where
\begin{align*}
	\tau_\nu&:= (\mathbf{1},i\vec{\sigma})
							\;,\quad
	\bar{\tau}_\nu := (\mathbf{1},-i\vec{\sigma})
							\;,\nonumber\\
	\gamma_\nu&:=\begin{pmatrix}
			0 & \tau_\nu	\\
			\bar{\tau}_\nu & 0
		  \end{pmatrix}
							\;,\nonumber\\
	\psi_i&:=\begin{pmatrix}
			\lambda_i	\nonumber\\
			\bar{\lambda}^i
		  \end{pmatrix}		\;,\;
	\bar{\psi}_i :=\psi^\dagger_i\gamma_0	\;,	\nonumber\\
	\tilde{M}_i&:=\begin{pmatrix}
		0 & -\tilde{\mu}_i\mathbf{1}_{2\times 2}\nonumber\\
		\tilde{\mu}_i\mathbf{1}_{2\times 2} & 0
		 \end{pmatrix}					\;.
\end{align*}

Finally, the
Euclidean Lagrangian density of the theory with those chemical potentials is
\cite{Gliozzi:1976qd,Brink:1976bc},
\begin{align}\label{FullLagrangian}
	\mathcal{L}_E&=\frac{1}{4}(F_{\mu\nu}^a)^2	\nonumber	\\
		&+ \frac{1}{2}(D_{\nu}X_p^a-i\mu_p\delta_{\nu,0} Y_p^a)^2
		+ \frac{1}{2}(D_{\nu}Y_p^a+i\mu_p\delta_{\nu,0} X_p^a)^2 \nonumber	\\
		&+ \frac{1}{2}\frac{1}{R^2}(\Phi_A^a)^2
		+\frac{1}{4}g^2(i[\Phi_A,\Phi_B]^a)^2 \nonumber \\
		&+\frac{1}{2}\bar{\psi}_i^a\Slash{D}\psi_i^a
		+\frac{1}{2}\bar{\psi}_i^a\tilde{M}_i\psi_i^a
		+\frac{1}{2}ig\bar{\psi}_i^a[(\alpha^p_{ij}X_p
					+i\beta^q_{ij}\gamma_5Y_q),\psi_j]^a  \;,
\end{align}
where $X_p$ and $Y_p$ are the real scalars defined by
$\phi_{1,p+1}=(1/\sqrt{2})(X_p+iY_p)$ (see (\ref{ComplexScalarRelations})),
the factor of $i$ in the quartic interaction term accounts for the
fact that a commutator of Hermitian fields is anti-Hermitian, and,
\begin{align*}
	F_{\mu\nu}^a&=\partial_{\mu}A^a_{\nu}-\partial_{\nu}A^a_{\mu}
			+ig[A_{\mu},A_{\nu}]^a	\;, \\
	D_{\nu}&=\partial_{\nu}+ig[A_{\nu},\,\cdot\,] \;, \\
	\Phi_A^a&=(X_1^a,Y_1^a,X_2^a,Y_2^a,X_3^a,Y_3^a) \;,
\end{align*}
with $a=1,\ldots ,N^2$, $A,B=1,\ldots ,6$, $p,q=1,\ldots,3$, and $i,j=1,\ldots,4$.
The $4\times 4$ matrices $\alpha^p$ and $\beta^q$ satisfy the relations,
\begin{equation}\label{abComm}
	\{\alpha^p,\alpha^q\}=-2\delta^{pq}\mathbf{1}_{4\times 4} \;,\quad	
	\{\beta^p,\beta^q\}=-2\delta^{pq}\mathbf{1}_{4\times 4} \;,\quad
	[\alpha^p,\beta^q]=0 \;.
\end{equation}
We take the convention where the color matrices are Hermitian.
The gauge and scalar fields are Hermitian fields and the components of
the fermions are complex Grassmann variables that are taking Hermitian
color matrix values.

We also have the relations,
\begin{equation*}
	[T^a,T^b]=if^{abc}T^c \;,\quad
	\{\gamma_\mu,\gamma_\nu\}=2\delta_{\mu\nu}\;,\quad
	\{\gamma_{\mu},\gamma_{5}\}=0 \;,\quad
	(\gamma_5)^2 = \mathbf{1}_{4\times 4}  \;,
\end{equation*}
where $T^a$ are the $N\times N$ Hermitian matrices of the $U(N)$ 
defining representation.
We normalize the basis of the $U(N)$ algebra so that,
\begin{equation*}
	\tr(T^aT^b)=\delta^{ab} \;.
\end{equation*}

In the present paper, we are interested in the half-BPS states
and therefore,
we turn on a single chemical potential,
\begin{equation*}
	\mu_p = \mu\delta_{p,1}	\;.
\end{equation*}
In what follows, we set the mass of the scalars, $1/R$, to 1 and
assume that all the energies, including the chemical potential,
$\mu$, are measured in the units of $1/R$.
\\

To construct the Wilsonian effective action of the theory
that is valid below the momentum scale of order $1=1/R$, we need
to integrate out the fields that are heavier than this scale.
Since the theory is on $S^3$, we expand the fields into their spherical
harmonics. Note that the topology of the sphere does not allow the zero
modes of the spatial vectors and spinors (they have the
lowest energies $1$ for the spatial vectors and $3/2$ for the spinors)
and the scalars have masses $1$ due to
the conformal scalar-curvature coupling. 
Now, the time component of the gauge field, $A_0$, is a massless scalar
and thus the only light mode of the theory
is the constant mode of $A_0$. This is the Lagrange multiplier
that enforces the Gauss' law constraint.

At first sight, therefore, we do not seem to have any dynamical field 
whose excitation
could be lower than the scale $1=1/R$. However, as one can see from the
Lagrangian, the chemical potential acts as negative mass terms for
$X$ and $Y$ scalars. Thus, those excitation energies could potentially be
smaller than the scale of the effective theory,
if the chemical potential is sufficiently
high, namely, $\mu\sim 1$ and we are going to assume such case.
This assumption no longer allows us to integrate out the zero modes of
$X,Y$ scalar fields.
Therefore, the resulting effective theory is the theory of constant
$X$ and $Y$ scalar fields on the time coordinate. This is a matrix quantum
mechanics.

The effective Lagrangian at the tree level is therefore given by,
\begin{align}\label{TreeLevelLagrangian}
	L&=\frac{1}{2}\tr\left[(D_0X-\mu Y)^2
				+(D_0Y+\mu X)^2
				-(X^2+Y^2)
				-g^2(i[X,Y])^2
						\right]
							\nonumber\\
	&=\tr\left[(D_0Z^\dagger+i\mu Z^\dagger)(D_0Z-i\mu Z)
				-Z^\dagger Z	
				-\frac{1}{2}g^2([Z,Z^\dagger])^2\right]
							\;,
\end{align}
where we Wick rotated to the Minkowski signature and defined the complex
fields $Z$ by $Z=(1/\sqrt{2})(X-iY)$.
\\

We now would like to obtain the one-loop effective potential of the
theory up to the order of $g^2$. 
This, in general, is a difficult task because the quartic
interaction term mixes the fields in the color space in a complicated way.
For this reason, we resort to
the special case where the scalar fields of the effective theory
are taking values in the flat
directions. This special case suffices to show what the loop corrections
generally do and might have some relevance to the recent development in the
theory with the commuting matrices \cite{Berenstein:2005aa,Berenstein:2005jq}.

The actual one-loop computation is rather involved and presented in
Appendix \ref{AppxOneLoop}. The method adopted there is the constant
background field method and the fields $(X,Y)$ are shifted to $(x+X,y+Y)$
where $(X,Y)$ are the background fields that are taking values in the
flat directions and $(x,y)$ are the arbitrary
fluctuations around them. By collecting the quadratic terms in
the shifted Lagrangian, the one-loop functional determinant is computed.
The result is,
\begin{equation}\label{OneLoopPtnl}
	V_1=Kg^2\sum_{m<n}|z_m-z_n|^2
					\;,
\end{equation}
where $z_m$ are the eigenvalues of the diagonal $Z$ matrix and $K$ is
defined to be,
\begin{equation*}
	K:=\frac{2}{\pi^2}
	\left\{-\frac{3}{4}+\ln2+C(\mu)+\frac{\mu^2}{4}
	(-2+2\gamma_E-2\ln2+\ln\Lambda^2)\right\}
						\;.
\end{equation*}
Here, $C(\mu)$ is a monotonically increasing function of $\mu$ and
ranges from $0.072$ at $\mu=0.8$ to $0.20$ at $\mu=1.2$. See Figure
\ref{CmuGraph} in Appendix \ref{AppxOneLoop}.
The constant
$\gamma_E$ is the Euler's constant and $\Lambda$ is an arbitrary renormalization
scale measured in the units of $1/R$. 
There is mass renormalization due to the non-vanishing
chemical potential and
the quantities $\gamma_E$, $\Lambda$ are associated with the
$\overline{MS}$-scheme that we adopted.


\section{Quantum Mechanics of the Theory}\label{QM}
In this section, we solve the resulting quantum mechanics without the 
interaction in two ways;
the operator method and the Schr\"odinger equation. Each method gives different
perspective on the decoupling limit that isolates
the half-BPS states from the other states of the full theory.
The former method shows that it is the low energy effective theory
of the scale $1=1/R$ and anti-particles of the theory are restricted to
the ground state. The latter shows that the system is the same
low energy effective theory but that is equivalent to the Landau
problem and the states are restricted to the lowest Landau level.

We treat the quartic interaction term in (\ref{TreeLevelLagrangian})
and the one-loop effect (\ref{OneLoopPtnl}) as perturbations to
the free theory. Thus, we first solve the free theory in
Subsection \ref{OpMethod}-\ref{ChangeMeasure}, and in Subsection
\ref{NoInteraction}, we show that 
the tree level quartic interaction has
no non-vanishing matrix elements. Finally in Subsection
\ref{LoopCorrection}, we discuss the matrix elements of the
one-loop potential (\ref{OneLoopPtnl}).


\subsection{Operator Method}\label{OpMethod}
We first solve the theory by using the creation and annihilation
operators. Although the physical set up and the motivations 
are somewhat different,
we would like to acknowledge the papers by Corley \textit{et al.}
\cite{Corley:2001zk}, Caldarelli and Silva \cite{Caldarelli:2004ig}
and especially by Takayama and Tsuchiya \cite{Takayama:2005yq}
that have carried out similar analyses to the one 
presented in this subsection.

We start the analysis from the complex field form of the
Lagrangian (\ref{TreeLevelLagrangian}) without the interaction term.
The conjugate momenta are,
\begin{align*}
	\Pi^a&:=\frac{\partial L}{\partial\dot{Z}^a}
		=\dot{Z}^{a*}+i\mu Z^{a*}
		-ig[A_0,Z^\dagger]^a
							\nonumber\\
	\Pi^{a*}&:=\frac{\partial L}{\partial\dot{Z}^{a*}}
		=\dot{Z}^a-i\mu Z^a
		-ig[A_0,Z]^a
							\;,
\end{align*}
where $a=1,\ldots,N^2$ is the color index.
The Hamiltonian is obtained through Legendre transformation.
Taking into account of the Hermiticity of the Hamiltonian,
we have,
\begin{align}\label{secondline}
	H&=\Pi^{a*}\dot{Z}^{a*}+\dot{Z}^a\Pi^a-L
						\nonumber\\
	&=\Pi^{a*}\Pi^a-i\mu \Pi^{a*}Z^{a*}+i\mu Z^a\Pi^a
					+Z^aZ^{a*}
							\\
	&\qquad -ig\left(\Pi^{a*}[A_0,Z^\dagger]^a+[A_0,Z]^a\Pi^a\right)
						\nonumber\\
	&=\tr\bigg[(\Pi^\dagger,Z)
				\begin{pmatrix}
					   1  & -i\mu \\
					i\mu &   1
				\end{pmatrix}
				\begin{pmatrix}
					\Pi \\
					Z^\dagger
				\end{pmatrix}
						\nonumber\\
	&\qquad\quad -ig\left(\Pi^\dagger[A_0,Z^\dagger]+[A_0,Z]\Pi\right)
						\bigg]
						\nonumber\;,
\end{align}
where the trace is taken over the color space.

The cyclic generalized coordinates $A_0$ are the constants of motion and their
equations of motion $\partial H/\partial A_0=0$ impose the Gauss'
law to the states. In Subsection \ref{SchurBasis}, we construct
the global $U(N)$ invariant states by hand and we are going to
ignore the $A_0$ part from now on.

Now define a unitary matrix,
\begin{equation*}
	U:=\frac{1}{\sqrt{2}}
			\begin{pmatrix}
				1 & -i	\\
				1 & i
			\end{pmatrix}\;,
\end{equation*}
which satisfies,
\begin{equation*}
	U
	\begin{pmatrix}
		1 & -i\mu  \\
		i\mu & 1
	\end{pmatrix}
	U^\dagger
	=
	\begin{pmatrix}
		1+\mu & 0  \\
		0 & 1-\mu
	\end{pmatrix}	\;.
\end{equation*}
Then, the Hamiltonian (without the $A_0$ factors) takes the form,
\begin{align}\label{abHamiltonian}
	H&=\tr\left[(\Pi^\dagger,Z)
				U^\dagger U
				\begin{pmatrix}
					   1  & -i\mu \\
					i\mu &   1
				\end{pmatrix}
				U^\dagger U
				\begin{pmatrix}
					\Pi \\
					Z^\dagger
				\end{pmatrix}
			\right]
							\nonumber\\
	&=\tr\left[(A^\dagger,B)
				\begin{pmatrix}
					   1+\mu  & 0 \\
					0 &   1-\mu
				\end{pmatrix}
				\begin{pmatrix}
					A \\
					B^\dagger
				\end{pmatrix}
			\right]
							\nonumber\\
	&=(1+\mu)\tr(A^\dagger A)
	+(1-\mu)\tr(BB^\dagger)	\;,
\end{align}
where we have defined,
\begin{equation}\label{ABDefinition}
	\begin{pmatrix}
		A\\
		B^\dagger
	\end{pmatrix}
	:=
	U
	\begin{pmatrix}
		\Pi\\
		Z^\dagger
	\end{pmatrix}
	=
	\frac{1}{\sqrt{2}}
	\begin{pmatrix}
		\Pi-iZ^\dagger	\\
		\Pi+iZ^\dagger
	\end{pmatrix}
							\;.
\end{equation}
It is clear from (\ref{abHamiltonian}) that we must have the
condition $|\mu|\leq 1$, otherwise the Hamiltonian is not bounded
from below. This condition is always true for free theories with
chemical potentials.

To quantize the system, we impose the canonical quantization
conditions,
\begin{equation*}
	[A^a,A^{b*}]=\delta^{ab}=[B^a,B^{b*}]	\;,\quad
				\text{otherwise commute}
							\;.
\end{equation*}
These are
equivalent to the conditions,
\begin{equation*}
	[Z^a,\Pi^b]=i\delta^{ab}	\;
	\Leftrightarrow			\;
	[Z^{a*},\Pi^{b*}]=i\delta^{ab}
					\;,\quad
				\text{otherwise commute}
							\;.
\end{equation*}
The commutation relations introduce zero point energy in
the Hamiltonian (\ref{abHamiltonian}) but we are going to
drop this by declaring the energy to be measured from
the ground state of the $A,B$ operators.

The authors of \cite{Corley:2001zk,Caldarelli:2004ig},
who have carried out similar analyses, diagonalize the matrix
corresponding to the $B$ operator at this point. However, this is
not valid. A matrix $M$ can be diagonalized by a unitary 
matrix if and only if $M$ is normal, \textit{i.e.},
$M$ satisfies $MM^\dagger=M^\dagger M$. The operator $B$ is not
normal because of the commutation relations
and hence one may not diagonalize it. The justification for
their procedure is more subtle and will be mentioned in
Subsection \ref{NoInteraction}.

Assuming $\mu>0$,
we see from (\ref{abHamiltonian})
that the operators $A^{a*}$ create the
quanta of unit energy $1+\mu$, \textit{i.e.}, the ones that
we are calling anti-particles in the original relativistic
theory
and $B^{a*}$ create particles
of energy quanta $1-\mu$.
However,
recall that we have constructed the effective action that is valid
below the scale $1=1/R$. Therefore, our system here cannot
afford to excite the anti-particles and the states of our theory
are always at the ground state of the $A$ operators.
Thus, our effective theory is a low energy effective theory of
the scale $1=1/R$ that constrains the anti-particles
to be at the ground state.

Obtaining the wavefunction is straightforward. From the definitions
of $A, B$ operators (\ref{ABDefinition}), we see that
the ground state is,
\begin{equation*}
	f_{0,0}=\prod_a\frac{1}{\sqrt{\pi}}\exp\left(-|Z^a|^2\right)	
							\;.
\end{equation*}
Then by applying $B^{a*}$ operators repeatedly,
we obtain,
\begin{equation*}
	f_{0,[n]}[Z^*]=\prod_a\frac{(Z^{a*})^{n_a}}{\sqrt{\pi n_a!}}
			\exp\left(-|Z^a|^2\right)	
							\;,
\end{equation*}
where $[n]$ and $[Z]$ are the shorthands for $(n_1,\ldots,n_{N^2})$ and
$(Z^1,\ldots,Z^{N^2})$, respectively. We note that the $n_i$ are
non-negative integers.
We thus have,
\begin{equation*}
	Hf_{0,[n]}[Z^*]=(1-\mu)\left(\sum_an_a\right)f_{0,[n]}[Z^*]
							\;,
\end{equation*}
and therefore this state evolves in time with the phase factor
$(1-\mu)\sum_an_a$.

Note that the set, $\{(Z^{a*})^{n_a}\}$,
forms a basis of holomorphic functions with respect to $[Z^*]$, so
in general, any wavefunction of the
theory can be written as,
\begin{equation*}
	\psi[Z^*]=f[Z^*]\exp\left(-\sum_a|Z^a|^2\right)
							\;,
\end{equation*}
where $f[Z^*]$ is a holomorphic function. We will take full advantage
of this fact in the formalism presented in the next subsection.

There is a problem with these solutions. Not all of them 
are satisfying
the Gauss' law and they contain
unphysical solutions. We must project the Hilbert space to the
physical subspace and we will deal with this issue in Subsection
\ref{SchurBasis}.
\\

Before we move on to the next method of solving the theory,
we would like to make comments about the special case
with $\mu=1$.
From the equation (\ref{secondline}), the Hamiltonian can be written as,
\begin{align*}
	H&=\tr(\Pi^\dagger\Pi+Z^\dagger Z)
		-\mu i\tr(\Pi^\dagger Z^\dagger -Z\Pi)
							\nonumber\\
	&=h-\mu j					\;,
\end{align*}
where we have defined,
\begin{equation*}
	h:=\tr(\Pi^\dagger\Pi+Z^\dagger Z)
							\;,\quad
	j:=i\tr(\Pi^\dagger Z^\dagger -Z\Pi)
							\;.
\end{equation*}
One notices that $h$ is the Hamiltonian of harmonic oscillators and
$j$ is the charge number operator.

When we have $\mu=1$, as one can see from (\ref{abHamiltonian}),
any particle states (as opposed to the anti-particle states) 
have the same energy. By dropping this constant from the Hamiltonian,
we have,
\begin{equation*}
	H|\psi\rangle =(h-j)|\psi\rangle =0
							\;,
\end{equation*}
for any state of the effective theory, $|\psi\rangle$, and
thus the physical states of our 
effective theory with
$\mu\sim 1$ purely consist of the half-BPS states. 
Moreover, since
$j$ is a conserved charge, $h$ and $j$, or equivalently, $H$
and $j$ commute. Therefore, there are infinitely many degenerate states
at the zero energy level and those states are labeled by the quantum
number of $j$. 
This is analogous to the system of the lowest Landau
level and in the next subsection, we show that it \textit{is} such 
a system.


\subsection{The Lowest Landau Level Projected Schr\"odinger Equation}
We solve the theory again, this time, 
by utilizing the lowest Landau level projection of
Girvin and Jach \cite{Girvin:1984}.
The first key observation is that the Hamiltonian of
the theory is the kinetic term that describes the Landau problem
and a potential. This can be seen by a simple rearrangement of
the Hamiltonian resulting from the tree level Lagrangian 
(\ref{TreeLevelLagrangian}).

From the real scalar expression of the Lagrangian, we have the
canonical momenta,
\begin{align*}
	P_X^a&=\frac{\partial L}{\partial\dot{X}^a}
		=\dot{X}^a-\mu Y^a
							\nonumber\\
	P_Y^a&=\frac{\partial L}{\partial\dot{Y}^a}
		=\dot{Y}^a+\mu X^a
							\;,
\end{align*}
where we discarded the $A_0$ fields as in the previous subsection.
The Legendre transformation leads to the Hamiltonian of the
the form,
\begin{align*}
	H&=P_X^a\dot{X}^a+P_Y^a\dot{Y}^a-L
							\nonumber\\
	&=\frac{1}{2}\tr\left[(P_X)^2+(P_Y)^2
				+2\mu (P_XY-P_YX)
				+(X^2+Y^2)\right]
							\nonumber\\
	&=\frac{1}{2}\tr\left[(P_X+\mu Y)^2
				+(P_Y-\mu X)^2
				+(1-\mu^2)(X^2+Y^2)\right]
								\;.
\end{align*}
The field redefinition in the Lagrangian,
$(X,Y)\rightarrow (X,Y)/\sqrt{c_1}$ with 
$c_1=2\mu/(1+\mu)^2$ and a trivial canonical transformation
$H\rightarrow c_2H$ with $c_2=(1+\mu)/2\mu$ can bring the Hamiltonian
into the form,
\begin{align*}
	H&=\frac{c_2}{2c_1}\tr\left[(P_X+c_1\mu Y)^2
				+(P_Y-c_1\mu X)^2\right]
				+\frac{1}{2}(1-\mu)\tr(X^2+Y^2)
							\nonumber\\
	&=H_0+V_0
							\;,
\end{align*}
where we have defined,
\begin{align}\label{Potential}
	H_0&:=\frac{c_2}{2c_1}\tr\left[(P_X+c_1\mu Y)^2
				+(P_Y-c_1\mu X)^2\right]	
								\;,
							\nonumber\\
	V_0&:=(1-\mu)\tr(Z^\dagger Z)
								\;.
\end{align}
In above, we redefined $Z^a:=(1/\sqrt{2})(X^a+iY^a)$.
The kinetic term of the Hamiltonian, $H_0$, exactly is the Hamiltonian
of the Landau problem%
\footnote{
Recall that a non-relativistic electrically charged particle
in a uniform magnetic
field is described by the Hamiltonian,
\begin{equation*}
	\tilde{H}=\frac{1}{2m}\left(\vec{p}+\frac{e}{c}\vec{A}\right)^2
							\;.
\end{equation*}
When one chooses the symmetric gauge,
\begin{equation*}
	\vec{A}=(yB/2,-xB/2,0)	\;,
\end{equation*}
which is related to the more common Landau gauge $\vec{A}=(0,-xB,0)$
by a gauge transformation,
the Hamiltonian becomes,
\begin{equation*}
	\tilde{H}=\frac{1}{2m}\left[(p_x+\frac{eB}{2c}y)^2
			+(p_y-\frac{eB}{2c}x)^2\right]	\;.
\end{equation*}
The energy of this system is quantized in the units of the cyclotron
frequency $eB/mc$.
}
and 
the Hamiltonian $H_0$ gives rise to the Landau level spacing
of $1+\mu$. As argued similarly in the previous subsection, 
any nonzero excitation
of the Landau level goes beyond the scale of our effective theory.
Therefore, our effective theory is described by the Hamiltonian
$H$ projected down to the lowest Landau level. It is interesting
to note that the projection to the ground state of the anti-particles
corresponds to the lowest Landau level projection.

The solutions to the Schr\"odinger equation
for $H_0$ is well-known (\textit{e.g.}, see \cite{Laughlin83})
and the eigenfunctions of the lowest Landau level are given by,
\begin{equation}\label{holobasis}
	f_{0,[n]}[Z]:=\prod_a\frac{(Z^a)^{n_a}}{\sqrt{\pi n!}}
		\exp\left({-|Z^a|^2}\right)
								\;,
\end{equation}
where $n_a$ are non-negative integers.
This is the same solution as the one obtained in
the previous subsection. However, this is the solution for the
kinetic part, $H_0$, of the full Hamiltonian and the eigenvalues
of $H_0$ for the lowest Landau level is zero. (More precisely,
the kinetic energy in this case 
is a constant of motion and we set the
constant to zero.) We must take into account of the potential
$V_0$.

As mentioned in the previous subsection, the set $\{(Z^a)^{n_a}\}$ forms
a basis of the space of holomorphic functions. Thus, any solutions
of the Hamiltonian $H$ projected down to the lowest Landau level
must have the form,
\begin{equation}\label{generalsoln}
	\psi[Z]=f[Z]\exp\left(-\sum_a|Z^a|^2\right)\;,
\end{equation}
where $f[Z]$ is a holomorphic function of $[Z]$.
Thus, all the wavefunctions of
the lowest Landau level are holomorphic functions of $[Z]$
times the exponential factor. This is a powerful constraint
on the form of wavefunctions and the systematic projection onto
the lowest Landau level had been developed by Girvin and
Jach \cite{Girvin:1984}.

We briefly summarize the prescription here. The important idea
involved is to absorb the non-holomorphic exponential factor
into the definition of the Hilbert space inner product. More
specifically, if $f[Z]$ and $g[Z]$ are the holomorphic parts of
wavefunctions, $\psi'$ and $\psi$, respectively,
then the inner product of the Hilbert space
is defined to be,
\begin{equation}\label{innerprod}
	(f,g):=\int d\mu[Z]f[Z]^*g[Z]	\;,
\end{equation}
where the integration measure $d\mu[Z]$ is given by,
\begin{equation}\label{originalmeasure}
	d\mu[Z]:=\prod_a\frac{1}{\pi}e^{-2|Z^a|^2}
		dX^adY^a	\;,
\end{equation}
such that the inner product of the wavefunctions is,
\begin{equation*}
	\langle\psi'|\psi\rangle=(f,g)	\;.
\end{equation*}
In this way, our Hilbert space now is the space of holomorphic
functions with the inner product defined as above.

The Hamiltonian of the system needs to be projected.
Consider a Hamiltonian of the form,
\begin{equation*}
	H[Z^*,Z]=H_0+V[Z^*,Z]	\;,
\end{equation*}
where $H_0$ is the Landau problem kinetic term and $V$ is
a potential. In general, the factors of $Z^{a*}$ in the potential
bring the states out of our Hilbert space and this is what
requires the projection. Girvin and Jach's prescription is to
``normal order'' the $Z^{a*}$ in the potential to the left of 
$Z^a$ and replace them as,
\begin{equation}\label{LLLP}
	Z^{a*}\rightarrow \frac{\partial}{\partial Z^a}\;.
\end{equation}
We have the factor of $2$ difference here from the Girvin and Jach
and this is due to the factor of $1/\sqrt{2}$ in
our definition of $Z$.
The projected potential, $\hat{V}$, satisfies the 
Schr\"odinger equation for the states in
the lowest Landau level,
\begin{equation}\label{LLLSchrodinger}
	\hat{V}f[Z]=Ef[Z]	\;.
\end{equation}

Thus, in our case with (\ref{Potential}), the projected potential is,
\begin{equation*}
	\hat{V}_0=(1-\mu)\sum_a\frac{\partial}{\partial Z^a}Z^a	
		=(1-\mu)\left(N^2+\sum_aZ^a\frac{\partial}{\partial Z^a}\right)
							\;.
\end{equation*}
The first term contributes to the zero point energy of the system
and as usual, we set that to zero.
Hence the differential equation that we must solve is,
\begin{equation*}
	(1-\mu)\sum_aZ^a\frac{\partial}{\partial Z^a}f[Z]=Ef[Z]
							\;,
\end{equation*}
and one can readily see that the solution is,
\begin{equation*}
	f_{[n]}[Z]=\prod_a\frac{(Z^a)^{n_a}}{\sqrt{n_a!}}
							\;,
\end{equation*}
with the eigenvalue $E_{[n]}=(1-\mu)\sum_an_a$. 
Recalling the fact that we absorbed
the non-holomorphic exponential factor (and $\pi$) 
into the Hilbert space
inner product, we in fact see that the eigenfunctions of $H_0$ 
are also the eigenfunctions of $\hat{V}_0$ with the eigenvalues 
$E_{[n]}$ as
above. This result is in agreement with the previous subsection.
But again, the solutions contain the ones that do not satisfy the
Gauss' law. In the next subsection, we project the Hilbert
space to the physical subspace.


\subsection{Orthogonal Basis of $U(N)$ Invariant States:
Schur Polynomials}\label{SchurBasis}
As described before, the solutions to the lowest Landau level
projected Schr\"odinger equation 
are holomorphic functions of $Z$ and therefore, the Hilbert space of
the theory is the space of holomorphic functions with the
inner product defined as (\ref{innerprod}). However, this Hilbert
space is too large and contains unphysical states.
Recall that we discarded the equations of motion for
$A_0$ in the previous subsections
by promising that we choose the states that satisfy the Gauss' law
as our physical
states. Therefore, the general holomorphic function is not the physical
solution to the theory at hand. The physical wavefunctions are the
holomorphic functions of $Z$ that is invariant under $U(N)$ conjugation.
A convenient basis for such holomorphic functions is the Schur
polynomials and the use of this basis was suggested by
Corley \textit{et al.} \cite{Corley:2001zk}. 
We are going to adopt this basis in our analysis
and will see the advantage of this.

The idea of the basis starts with the observation that general half-BPS
operators of a fixed $R$-charge, $n$, have the form,
\begin{equation*}
	\{\tr(Z^{l_1})\}^{k_1}\cdots\{\tr(Z^{l_m})\}^{k_m}
						\;,
\end{equation*}
where the integers $l_i$ and $k_i$ satisfy,
\begin{equation*}
	n=\sum_{i=1}^ml_ik_i		\;.
\end{equation*}
Schur polynomials for the complex matrix are linear combinations of
such polynomials and they are given by,
\begin{equation*}
	\chi_{n,R}(Z)=\frac{1}{n!}\sum_{\sigma\in S_n}
			\chi_R(\sigma)\tr(\sigma Z)
							\;,
\end{equation*}
where $S_n$ is the symmetric group, $R$ is its representation and
$\chi_R(\sigma)$ is the character of $\sigma$ in $S_n$. The last
factor can be explicitly written as,
\begin{equation*}
	\tr(\sigma Z)=\sum_{i_1,\ldots,i_n}Z_{i_1i_{\sigma(1)}}
	\cdots Z_{i_ni_{\sigma(n)}}	\;,
\end{equation*}
where we packaged the $N^2$ complex degrees of freedom $Z^a$
into a $N\times N$ complex matrix, that is, each degree of freedom 
now is given
by $Z_{ij}$ which is defined by $Z_{ij}=\sum_aZ^a(T^a)_{ij}$.
For example, when $n=2$, there are symmetric and antisymmetric
representations for the symmetric group and
the Schur polynomials in this case are,
\begin{equation}\label{n2Example}
	\chi_{2,S}(Z)=\frac{1}{2}\{(\tr Z)^2+\tr(Z^2)\}
						\;,\quad
	\chi_{2,A}(Z)=\frac{1}{2}\{(\tr Z)^2-\tr(Z^2)\}
						\;.
\end{equation}

The virtue of this basis is that they are orthogonal with respect to
the inner product defined in (\ref{innerprod}),
\begin{equation*}
	(\chi_{n_1,R_1}(Z),\chi_{n_2,R_2}(Z))\sim\delta_{n_1,n_2}\delta_{R_1,R_2}
							\;,
\end{equation*}
and spans the space of $U(N)$ invariant functions 
of the complex matrix $Z$.
The proof of Corley \textit{et al.} \cite{Corley:2001zk} for the
orthogonality is valid
here with the only modification being the expectation value of the
operators is replaced by the inner product (\ref{innerprod}).
The modification does not change the proof by noting that we have,
\begin{equation*}
	((Z_{ij})^m,(Z_{kl})^n)=m!\delta_{i,k}\delta_{j,l}\delta^{m,n}
								\;.
\end{equation*}

It is important to note that the basic building block of 
the Schur polynomials
$\{\tr(Z^m)\}^n$ can be reduced to a polynomial of the
eigenvalues. In fact, given a complex matrix $Z$, there
exists a unitary matrix $U$ such that,
\begin{equation}\label{triangulation}
	Z=UWU^\dagger	\;,
\end{equation}
where $W$ is a triangular complex matrix, \textit{i.e.}, we have
$W_{ij}=0$ for all $i>j$ and the diagonal elements $z_i:=W_{ii}$
are the eigenvalues. Thus we have,
\begin{equation*}
	\{\tr(Z^m)\}^n=\{\tr(W^m)\}^n
	=\left(\sum_{i=1}^Nz_i^m\right)^n
						\;,
\end{equation*}
as claimed. This shows that the physical wavefunctions are 
functions only of $[z]:=(z_1,\ldots,z_N)$ and do not
involve the off-diagonal elements of the matrix $W$.
\\

Let us consider the eigenvalues of the Schur polynomials with respect
to the potential $V_0$ as in (\ref{Potential}).
Since we have,
\begin{equation*}
	\tr(Z^\dagger Z)=\sum_i|z_i|^2
			+\sum_{l<m}|W_{lm}|^2
							\;,
\end{equation*}
the projected potential is,
\begin{equation*}
	\hat{V}_0=(1-\mu)\left(\sum_iz_i\frac{\partial}{\partial z_i}
		+\sum_{l<m}W_{lm}\frac{\partial}{\partial W_{lm}}\right)
							\;,
\end{equation*}
where we again dropped the constant.
For a polynomial of the form,
\begin{equation*}
	\left(\sum_iz_i^{l_1}\right)^{k_1}
		\cdots\left(\sum_iz_i^{l_m}\right)^{k_m}
						\;,
\end{equation*}
where the integers $l_i$ and $k_i$ satisfy,
\begin{equation*}
	n=\sum_{i=1}^ml_ik_i		\;,
\end{equation*}
it is not hard to see that this is an eigenfunction of the operator
$\sum_iz_i(\partial/\partial z_i)$ with the eigenvalue $n$.
Thus, by construction, the Schur polynomial $\chi_{n,R}[z]$ also
is an eigenfunction of the operator with the same eigenvalue.
Hence we have,
\begin{equation}\label{Solution1}
	\hat{V}_0\chi_{n,R}[z]=(1-\mu)n\chi_{n,R}[z]
							\;.
\end{equation}
This is the physical solution to the lowest Landau level projected 
Schr\"odinger
equation. But the story is not complete. The triangulation of the
matrix $Z$ in (\ref{triangulation}) causes the change in the
inner product measure $d\mu[Z]$ of (\ref{originalmeasure}) and
we discuss the change of the measure in the next subsection.


\subsection{Change of the Measure and the Solution}\label{ChangeMeasure}
In this subsection, we reduce the complex matrix quantum mechanics
into the quantum mechanics of the eigenvalues. This procedure is
described in Chapter 15 and Appendix 33 of the book by Mehta
\cite{Mehta} and also reproduced in the paper by Takayama and
Tsuchiya \cite{Takayama:2005yq}.

We first rewrite the
inner product measure of the Hilbert space (\ref{originalmeasure})
as,
\begin{equation*}
	d\mu[Z]=\prod_{i,j}d\opname{Re}Z_{ij}d\opname{Im}Z_{ij}
			\exp\left(-2\tr(Z^\dagger Z)\right)
						\;.
\end{equation*}
We dropped the irrelevant overall constant that comes from the
Jacobian associated with the change of the variables from
$X^a$ and $Y^a$ to $\opname{Re}Z_{ij}$ and $\opname{Im}Z_{ij}$.

Now, under the change of the variables (\ref{triangulation}),
a straightforward computation for the
variation of $Z$ shows that the measure takes the form,
\begin{align}\label{ReducedMeasure}
	d\mu[Z]=\prod_{i<j}|z_i-z_j|^2\prod_kdz_kdz_k^*
	&\prod_{l<m}dW_{lm}dW_{lm}^*dM_{lm}dM_{lm}^*
							\nonumber\\
		&\times\exp\left(-2\sum_i|z_i|^2
			-2\sum_{j<k}|W_{jk}|^2\right)	\;,
\end{align}
where $dM$ is the Hermitian matrix
defined by $dM=-iU^\dagger dU$ with the constraints
$dM_{ii}=0$. The latter constraints are possible due to the fact that 
the unitary matrix $U$ that triangulates $Z$ is not unique.

Since our potential and wavefunctions do not involve $M$, these
degrees of freedom integrate to a constant. Also in the previous
subsection, we saw that $W$ sector of the potential had vanishing
matrix element due to the fact that the wavefunctions are independent
of $W$. Therefore, the factors of $M$ and $W$ may be dropped from
the measure. We will further elaborate on this point in the
next subsection.

Note that the first piece in the measure is the conjugate squared
Van der Monde determinant $\Delta:=\prod_{i<j}(z_i-z_j)$. It is customary
to absorb this factor into a wavefunction of the theory by,
\begin{equation}\label{FermiWavefunction}
	\psi_F[z]:=\Delta\psi[z]	\;,
\end{equation}
and at the same time, modify the Hamiltonian by,
\begin{equation}\label{FermiHamiltonian}
	H_F:=\Delta H\frac{1}{\Delta}	\;.
\end{equation}
Under these redefinitions, the Van der Monde determinant factor
in the measure (\ref{ReducedMeasure}) must be omitted. By dropping
the factors of $M$ and $W$, we now have the reduced measure,
\begin{equation}\label{NewMeasure}
	d\hat{\mu}[z]=\prod_i\frac{1}{\pi}dz_idz_i^*
		\exp\left(-2|z_i|^2\right)
								\;,
\end{equation}
where we reinserted the normalization factor of $\pi$.
Since $\Delta$ is anti-symmetric under the exchange of any
$z_i$, the wavefunction $\psi_F$ is fermionic, hence our theory
boils down to the system of identical fermions and this is
\textit{analogous} to the system of electrons under a strong uniform
magnetic field.

Given the redefinition (\ref{FermiWavefunction}) and the solution
of the previous subsection (\ref{Solution1}), we now have the
final time-dependent physical solution,
\begin{equation*}
	\psi_F[z,t]=\sum_{n,R}C_{n,R}e^{-i(1-\mu)nt}\Delta\chi_{n,R}[z]
							\;,
\end{equation*}
where $C_{n,R}$ are constants
and the inner product of the Hilbert space is given by the measure
$d\hat{\mu}[z]$ of (\ref{NewMeasure}).

We see from the solution that the degenerate 
states of the lowest Landau level
in $\mu=1$ case is lifted for $\mu\neq1$.
The ground state is the lowest filling
lowest Landau level where the $N$ states are filled from the state
with the lowest angular momentum to the 
higher without any gap and the excited
states are the state filling with gaps. The energies of the states
are proportional to the $R$-charges, $n$, and the states are responding
to the chemical potential, which is lower than the mass of the scalars,
according to the size of the $R$-charges that they have.

However, the degeneracy is not
completely lifted. At the energy level $n$, the number of degenerate
states is the number of $S_n$ irreducible representations, that is,
the number of Young diagrams for $n$ boxes.
This
is also the number of the ways to partition $n$ and as $n$ becomes
large, this number grows exponentially. This is the Hagedorn behavior.
We note that $n$ may not be too large that the energy goes
beyond the scale of the effective theory, namely, we must have
the condition,
\begin{equation}\label{ValidityCondition}
	n<\frac{1}{1-\mu}
					\;.
\end{equation}
Thus, in this effective theory,
such Hagedorn behavior occurs only when $N$ is large and 
the chemical potential
is slightly below the mass of the scalars.

We also note that when $\mu\neq1$, the states acquire relative
phase differences along with the time evolution. This is not the
case for $\mu=1$ where all the states evolve with the same
phase factor and they do not have relative phases.


\subsection{The Tree Level Interaction Term}\label{NoInteraction}
Recall that the measure (\ref{ReducedMeasure}) contains the
factors of $M$ and $W$, and we argued that these factors
may be dropped. In this subsection, we elaborate on this point
and show that the quartic interaction in (\ref{TreeLevelLagrangian})
does not play a role at the tree level.

First, observe that any potential that is invariant under
$U(N)$ conjugation does not contain
the factor of $M$ because the triangulation matrix $U$ in
(\ref{triangulation}) must drop out. Moreover, the physical
wavefunctions do not contain the factor of $M$ either, so 
in general, this factor integrates to a constant.

Second, again, since the wavefunctions do not contain the
factors of $W_{lm}$, non-vanishing matrix elements
of the potential that contain $W_{lm}$ must have the form
$|W_{lm}|^{2k}$ where $k$ is an integer. However, this form of
the potential does not occur under the lowest Landau level
projection. The prescription for the projection instructs us
to replace $W_{lm}^*$ to the derivatives $\partial/\partial W_{lm}$,
and therefore, apart from the zero point energy,
the matrix elements of the potential that
contain only $W$ are always zero with respect to the basis
discussed in Subsection \ref{SchurBasis}.
This means that we may begin with
the potential whose $W$ factors, that are not mixing with
$z_i$, are all set to zero. In this very subtle sense,
the incorrect diagonalization of $B$ mentioned in 
Subsection \ref{OpMethod} is ``valid.''

When the elements of $W$ are mixing with $z$, the situation
is different. For example, the term of the form $|W|^{2l}|z|^{2n}$
would give the contribution $(N(N-1)/2)^{2l}|z|^{2n}$.
The quartic interaction in (\ref{TreeLevelLagrangian}) indeed
contains the $W$-$z$ mixing components. However, we are going to
show that they do not contribute to the matrix elements of
the Hamiltonian. The quartic term can be written as,
\begin{equation*}
	\frac{1}{2}\tr([Z,Z^\dagger]^2)=\frac{1}{2}\tr([W,W^\dagger]^2)=
	W^*_{ji}W^*_{lk}W_{jk}W_{li}-W^*_{ji}W^*_{kj}W_{kl}W_{li}
						\;,
\end{equation*}
where we ``normal ordered'' the $W^*$ according to the lowest Landau
level projection prescription.
After replacing the $W^*$ to the derivatives $\partial/\partial W$,
one may try to
bring the derivatives toward right. In this process, each term gives
a constant, the terms of the form $W(\partial/\partial W)$
and the terms with the form $WW(\partial^2/\partial W^2)$ where the
indices are assumed to be contracted properly. The constants from the
two terms cancel. Now, because the wavefunctions are functions only
of $W_{ii}=z_i$, the terms of the form $W(\partial/\partial W)$
in the both terms only have non-vanishing elements with
$W_{ii}(\partial/\partial W_{ii})$ and hence they cancel.
Similar argument applies to the terms of the form 
$WW(\partial^2/\partial W^2)$. Thus, at the tree level,
the interaction does not play a role and the absence of it implies
that the states of the theory do not mix.

One might have expected this result from the usual behavior of the
half-BPS states. However, it is nice to see that this works in terms
of the lowest Landau level projection. And we point out that
the nonrenormalization theorem is not valid for the effective
theory with the chemical potential. In fact, we have the one-loop
corrections to the effective potential as in (\ref{OneLoopPtnl}).
In the next subsection, we are going to discuss the perturbation
theory of this one-loop potential.


\subsection{The One-Loop Perturbation}\label{LoopCorrection}
We are going to treat the one-loop potential obtained in
(\ref{OneLoopPtnl}) as a perturbation. 
We remind the reader that the one-loop correction to the effective
action was obtained by restricting the fields $X,Y$ to take the
values only in the flat directions. This is rather a special case,
however, we will see the general behavior of the loop corrections
from this case.

The lowest Landau level projected form of the potential is,
\begin{equation*}
	\hat{V}_1=Kg^2\sum_{m<n}\left(
		\frac{\partial}{\partial z_m}
		-\frac{\partial}{\partial z_n}
				\right)
		(z_m-z_n)
						\;.
\end{equation*}
First, note that the perturbation does not change the order of
polynomials on which it acts, implying that the potential
does not have non-vanishing matrix elements that connect states
with different charges. This is a general statement for any
corrections to the effective action because of the gauge invariance.
This is consistent with the fact that the $R$-charge is a conserved
quantity.

However, we expect the perturbation to mix the states with the same
$R$-charge and lifts the degeneracy that we observed in the end of 
Subsection \ref{ChangeMeasure}.
To illustrate this, we consider a concrete system with
the number of the color $N=3$, and the states with $R$-charge $n=2$.
The states are given in (\ref{n2Example}) for $n=2$ and they can
be explicitly written as,
\begin{equation*}
	\chi_{2,S}=\frac{1}{24}\left\{(\sum_{i=1}^3z_i)^2+\sum_{i=1}^3z_i^2\right\}
							\;,\quad
	\chi_{2,A}=\frac{1}{12\sqrt{2}}\left\{(\sum_{i=1}^3z_i)^2-\sum_{i=1}^3z_i^2\right\}
							\;,
\end{equation*}
where we normalized the states with respect to $|\Delta|^2d\hat{\mu}[z]$
which was discussed in Subsection \ref{ChangeMeasure}.
Then, we have,
\begin{align*}
	\hat{V}_1\chi_{2,S}&=Kg^2(8\chi_{2,S}-2\sqrt{2}\chi_{2,A})
							\;,\nonumber\\
	\hat{V}_1\chi_{2,A}&=Kg^2(-2\sqrt{2}\chi_{2,S}+10\chi_{2,A})
							\;,
\end{align*}
and in the basis of $(\chi_{2,S},\chi_{2,A})^T$, we have the
matrix expression,
\begin{equation*}
	(\chi_{2,R'},\hat{V}_1\chi_{2,R})
		=Kg^2
		\begin{pmatrix}
			8 & -2\sqrt{2} \\
			-2\sqrt{2}&  10
		\end{pmatrix}
					\;.
\end{equation*}
Given this, one can solve the degenerate perturbation
theory up to $g^2$.
The energy eigenstates and their energies are,
\begin{align*}
	&\frac{1}{\sqrt{3}}(\sqrt{2}\chi_{2,S}+\chi_{2,A})
						\;,\qquad
	E=2(1-\mu)+6Kg^2
						\;,\nonumber\\
	&\frac{1}{\sqrt{3}}(-\chi_{2,S}+\sqrt{2}\chi_{2,A})
						\;,\quad
	E=2(1-\mu)+12Kg^2	
						\;.
\end{align*}
We see here the mixing of the states and the splitting in
the energy level.

Even though we worked out a simple example, we should expect
the mixing and the energy level splitting for any corrections
to the effective action. We would like to emphasize that we
do not know how the theory behaves at the strong coupling
region.


\section{AdS/CFT Interpretation/Speculation}\label{Interpretation}
As described in Subsection \ref{OpMethod}, at the value of
the chemical potential $\mu\sim 1$, the states of the effective theory
are the half-BPS states, $|\psi\rangle$, satisfying
$(h-j)|\psi\rangle=0$. In AdS/CFT correspondence, those states are
conjectured to be the various gravitons in the dual gravity theory.
When the $R$-charge
is small compared to $N$, those states correspond to the Kaluza-Klein
gravity modes propagating in the bulk.
When the $R$-charge is comparable to the large number
of the color $N$, the Schur polynomial of totally symmetric representation
corresponds to the giant graviton in AdS bulk while totally antisymmetric
case corresponds to the giant graviton in $S^5$. (See the papers by
Balasubramanian \textit{et al.} \cite{Balasubramanian:2001nh} and
Corley \textit{et al.} \cite{Corley:2001zk}.)
And as mentioned in Introduction, all the geometries that correspond 
to the half-BPS states of $\mathcal{N}=4$ SYM are constructed by LLM
in \cite{Lin:2004nb}. In the effective theory that we have considered,
the conjectured dual of those gravity systems are all degenerate
at $\mu=1$ and they evolve in time without phase differences.

Now let us discuss the case with $\mu\sim 1$ but $\mu<1$.
Even though our analysis
has been carried out for general $N$,
in order for the gravity
calculation to be valid, we must take the large $N$ limit. Since
the giant graviton states correspond to the particular representations
of $S_n$ when $n$ is comparable to $N$, the condition 
(\ref{ValidityCondition}) tells us that this picture is valid only
when the chemical potential is slightly lower than the mass of
the scalars. More specifically, we must have $N\lesssim 1/(1-\mu)$.
Put it another way, 
the excitations that correspond to the giant gravitons
cost too much in the low energy effective theory and they are not excited.
The gravitons with $R$-charges that satisfy (\ref{ValidityCondition})
are allowed to appear and they respond to the chemical
potential proportional to their charges as expected.

Note that the ground state is the lowest filling lowest Landau level,
and according to LLM \cite{Lin:2004nb}, this state
is conjectured to be the geometry, AdS$_5\times S^5$. Thus, when
the chemical potential is lower than the mass of the scalars, 
AdS$_5\times S^5$ geometry acquires a special status as the
ground state of quantum mechanically superposed other geometries.
However, because of the $R$-charge conservation, the theory does
not have non-vanishing matrix elements that connect the states
with different $R$-charges and in this sense, the ground state is
not very special.

So far in this section, we have been discussing the correspondence
between the field theory and the gravity pictures as if the
identification is proved. The author, however, feels that this
is too naive especially
for the theory with the chemical
potential. In our effective theory, the half-BPS states are not
protected against the change in the coupling constant, and as
we mentioned at the end of Subsection \ref{LoopCorrection},
we do not know how the theory behaves at the strong coupling
region where the gravity picture is supposed to be valid.
The discussions above should be treated as speculations
rather than interpretations.

In the work of LLM \cite{Lin:2004nb}, the geometry
is obtained by specifying 
the boundary condition of a particular function on a
two dimensional plane in ten dimensions. It is conjectured that this
two dimensional plane corresponds to the phase space of the free
fermions that arises from the reduction of the SYM on $\mathbb{R}\times
S^3$ onto $\mathbb{R}$ (\textit{c.f.} Berenstien's work
\cite{Berenstein:2004kk}).
Notice that the phase space coordinates
do not commute quantum mechanically and the question was raised;
in what sense the two coordinates become non-commutative? 
As already noted by Berenstein in \cite{Berenstein:2004kk},
this non-commutativity seems to arise from the familiar
feature of the lowest Landau level.
In our
analysis above, the variable $z^*$ is replaced by the derivative
$\partial/\partial z$ under the lowest Landau level projection
(\ref{LLLP}) and hence
the commuting variables $z,z^*$ become non-commutative.
In fact, we have $[z,-i\partial/\partial z]=i$. Let us rewrite
this in terms of the real variables defined by 
$x_1=(1/i\sqrt{2})(z-z^*)$
and $x_2=(1/\sqrt{2})(z+z^*)$. Upon the lowest Landau level
projection, these variables become 
$\hat{x}_1=(1/i\sqrt{2})(z-\partial/\partial z)$ and 
$\hat{x}_2=(1/\sqrt{2})(z+\partial/\partial z)$, and they 
satisfy the commutation relation $[\hat{x}_1,\hat{x}_2]=i$.
This appears to
be the relation of conjugate variables that form two dimensional
phase space. 

In the decoupling limit of Maldacena \cite{Maldacena:1997re},
the low energy field theory living on a stack of $N$ D3-branes in
Type IIB string theory is $\mathcal{N}=4$ SYM and the transverse
coordinates of the D3-branes are the scalar fields of the SYM.
If we literally carry over this identification to our SYM on
$\mathbb{R}\times S^3$, then the eigenvalues $z_i$ are the
coordinates of the D3-branes. Thus, projecting the (low energy
effective) theory down to the lowest Landau level, the positions
of the D3-branes no longer commute, again, if we boldly carry
over the analogy. Also this seems to suggest that the phase
space distribution (known as the droplet configuration) 
of LLM is the distribution of the D3-branes.
In fact, the total area of the phase space configuration is 
calculated to be $N$ by LLM. However, the author
of the present paper is not aware of such identification
and the proof.

\begin{figure}
\centerline{\scalebox{.65}{\includegraphics{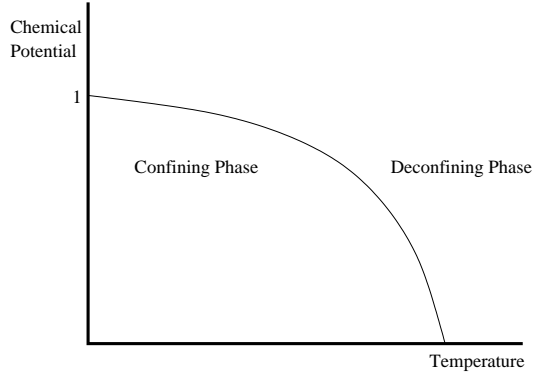}}}
\caption{\footnotesize Schematic drawing of the phase diagram 
of the finite temperature
free SYM on sphere with the $R$-symmetry chemical potential.}
\label{PhaseDiagram}
\end{figure}

The analysis carried out in the present work has another aspect
in AdS/CFT correspondence. In \cite{YY:2005}, the effective
theory of the finite temperature 
free SYM on $S^1\times S^3$ with the $R$-symmetry
chemical potentials has been considered. There, all the massive
modes, including the scalars with chemical potentials, are
integrated out and the effective theory becomes the matrix model
of the constant $A_0$ mode that implements the Gauss' law constraint.
The partition function is derived for the theory and a phase
structure of the theory in the grand canonical ensemble has been
found. The schematic phase diagram of the theory is shown in
Figure \ref{PhaseDiagram}.
As indicated in the diagram, the phases are named as ``confining
and deconfining'' phases. This is because the
Polyakov loop and the large $N$ scaling behaviors are the same
as the ones observed in the flat space gauge theories. (It still
is too naive to directly identify those phases to the confining/deconfining
phases. For example, the real QCD does not confine at zero coupling.)

The AdS/CFT correspondence conjectures that the confining and 
deconfining phases are dual to the AdS space with string gas
and AdS space with a black hole, respectively \cite{Witten:1998qj,Witten:1998zw}.
When one turns on the chemical potential, the conjecture reads that
the
confining phase as the dual of the AdS space with charged gas 
and the deconfining phase as the Reissner-Nordstr\"om-AdS black hole.
Now recall that we have worked out zero temperature effective
field theory with the chemical potential less than or equal to
$1$. 
If the states that we obtained in the field theory side 
correspond to the geometries of LLM, then they are
smooth and do not have horizons. When the back reactions of the
gravitons can be ignored, which we think is plausible when the
$R$-charge is small compared to $N$,
those geometries are gravitons
in AdS space. 
This seems to resemble the conjectured gravity
picture of the confining phase in the field theory side and
fits to the zero temperature axis of Figure \ref{PhaseDiagram}
below the chemical potential $1$.


\section{Outlook}\label{Outlook}
We have carried out the analysis for the effective action at the
tree level and the one-loop level with the commuting fields.
We also turned on one out of three possible chemical potentials.
It is natural to extend the
analysis to two or three nonzero chemical potentials and also
construct the general 
higher loop corrections to the effective action.

For the case with more than one nonzero chemical potentials,
we have more than one complex fields in the effective theory and
one may not
be able to diagonalize or triangulate those matrices simultaneously.
We expect the theory to become very complicated to analyze.
If we assume that the half-BPS states do not receive corrections
from the interactions as in the flat space case, it is plausible
to consider the theory of the commuting matrices so that the
quartic interactions do not contribute. 
There are arguments that at the classical level, the matrices
must be commuting and the works on such commuting
matrices have been carried out by Berenstein \textit{et al.}
\cite{Berenstein:2005aa,Berenstein:2005jq}.
Some investigations need to be done to see how this scenario
works in the effective theory with chemical potentials.

The higher loop calculations are worth
doing because the loop corrections to the effective action
is crucial to answer the question; what happens to the theory
when $\mu>1$? We could not investigate this issue in the present
work, because the one-loop computation was limited to the 
commuting fields.
The theory may go through a phase transition as
found at the high temperature end \cite{YY:2005}. We should
expect that the phase structure of a theory to be robust
against the change in the coupling, so an analysis at the
weak coupling has more hope of capturing the dynamics at
the strong coupling region. This is currently under investigation.

We have considered the theory on $\mathbb{R}\times S^3$ and this is
a zero temperature field theory on sphere. 
As mentioned at the end of the previous section,
there is a finite
temperature version of the theory on $S^1\times S^3$ where the
circumference of $S^1$ is the inverse of the temperature. 
Witten pointed out in \cite{Witten:1998zw} that
the conformal
symmetry requires the parameter of the theory be the ratio of
the two radii, $TR$, where $T$ is the temperature and $R$ is the
radius of $S^3$. Thus, the low and high temperature limits are 
equivalent to the theory on a small sphere and on a nearly flat
three space, respectively. The high temperature end of the
theory with chemical potentials has been carried out in
\cite{YY:2005}. Thus the natural next step is to consider
the low temperature limit of the theory. This is also 
under investigation.

\section*{Acknowledgments}
I would like to thank A.~Karch, A.~Kryjevski and L.~Yaffe 
for discussions. I also would like to thank A.~Karch and S.~Yoon
for reading through the drafts of this work.
This work was partially supported by the DOE under contract
DE-FG03-96-ER-40956.


\appendix
\section{The One-Loop Effective Potential}\label{AppxOneLoop}
In this appendix, we derive the one-loop effective potential 
(\ref{OneLoopPtnl}) of the theory. 
We restrict ourselves to the
the special case where the matrix valued fields of
interest are taking values in the flat directions and they
are diagonal. We adopt the background
field method. The $X,Y$ fields of the theory in the Lagrangian
(\ref{FullLagrangian})
are shifted to
$x+X,y+Y$ where $X,Y$ are now the diagonal constant backgrounds
and $x,y$ are the arbitrary fluctuations around them. We denote
the diagonal components of $X,Y$ by $X_n,Y_n$.
The shift causes the
quadratic terms that mix the gauge fields and the scalars. To get rid
of the mixed terms as much as possible, we employ the $R_\xi$-gauge,
\begin{equation*}
	\frac{1}{2\xi}\left\{\partial_\mu A_\mu-
			i\xi g\left([x,X]+[y,Y]\right)\right\}^2
							\;,
\end{equation*}
with $\xi=1$.

The quadratic terms in the shifted (Euclidean) Lagrangian with the
above choice of the gauge are,
\begin{align}\label{QuadraticTerms}
	\sum_{m<n}\bigg[
	&(A_\iota)_{mn}^*(-\partial^2+g^2M_{mn}^2)(A_\iota)_{mn}
							\nonumber\\
	+&(\Phi_A)_{mn}^*(-\partial^2+1+g^2M_{mn}^2)(\Phi_A)_{mn}
							\nonumber\\
	+&\left((A_0)_{mn}^*,x_{mn}^*,y_{mn}^*\right)\mathcal{M}
		\begin{pmatrix}
			(A_0)_{mn}
				\\
			x_{mn}
				\\
			y_{mn}
		\end{pmatrix}
							\nonumber\\
	+&(\bar{\psi}_i)_{mn}^*
		\{\delta_{ij}(\Slash{\partial}+\tilde{M}_i)
		+ig(\alpha^1_{ij}X_{mn}+i\beta^1_{ij}\gamma_5Y_{mn})\}
		(\psi_j)_{mn}
				\bigg]
							\;,
\end{align}
where $m,n$ are the color indices, the index $\iota$ runs 
from $1$ to $3$ for the spatial components
of the gauge fields, the index $A$ on the scalar fields
$\Phi$ runs from 3 to 6 excluding the $x,y$ scalars,
and we have defined $X_{mn}:=X_m-X_n$, $\;Y_{mn}:=Y_m-Y_n$, 
$\;M_{mn}^2:=X_{mn}^2+Y_{mn}^2$ and a matrix,
\begin{equation*}
	\mathcal{M}:=
		\begin{pmatrix}
			-\partial^2+g^2M_{mn}^2 &
			-2\mu gY_{mn}	&
			2\mu gX_{mn}	\\

			2\mu gY_{mn}	&
			-\partial^2+1-\mu^2+g^2M_{mn}^2 &
			2i\mu\partial_0	\\

			-2\mu gX_{mn}	&
			-2i\mu\partial_0	&
			-\partial^2+1-\mu^2+g^2M_{mn}^2
		\end{pmatrix}
					\;.
\end{equation*}
For the explicit computation of the fermion sector, one may use
the matrices,
\begin{align*}
	\alpha^1&=
	\begin{pmatrix}
		0 & \sigma_1	\\
		-\sigma_1 & 0
	\end{pmatrix}	\;,\;
	\alpha^2=
	\begin{pmatrix}
		0 & -\sigma_3	\\
		\sigma_3 & 0
	\end{pmatrix}	\;,\;
	\alpha^3=
	\begin{pmatrix}
		i\sigma_2 & 0	\\
		 0 & i\sigma_2
	\end{pmatrix}	\;,\;
				\nonumber\\
	\beta^1&=
	\begin{pmatrix}
		0 & i\sigma_2	\\
		i\sigma_2 & 0
	\end{pmatrix}	\;,\;
	\beta^2=
	\begin{pmatrix}
		0 & \sigma_0	\\
		-\sigma_0 & 0
	\end{pmatrix}	\;,\;
	\beta^3=
	\begin{pmatrix}
		-i\sigma_2 & 0	\\
		 0 & i\sigma_2
	\end{pmatrix}	\;,
\end{align*}
that satisfy the relations (\ref{abComm}).

Now the logarithms of the functional determinants that result from
(\ref{QuadraticTerms}) are,
\begin{align}\label{TraceLog}
	&-3\tr\ln(\omega_0^2+\Delta_g^2+g^2M_{mn}^2)
	-4\tr\ln(\omega_0^2+\Delta_s^2+g^2M_{mn}^2)
							\nonumber\\
	&-\tr\ln(\omega_0^2+\Delta_{A_0}^2+g^2M_{mn}^2)
							\nonumber\\
	&-\tr'\ln\left[\{\omega_0^2+(\sqrt{\Delta_s^2+g^2M_{mn}^2}-\mu)^2\}
		\{\omega_0^2+(\sqrt{\Delta_s^2+g^2M_{mn}^2}+\mu)^2\}
		\right]
								\\
	&+4\tr\ln\left[\{\omega_0^2+(\sqrt{\Delta_f^2+g^2M_{mn}^2}-\mu/2)^2\}
		\{\omega_0^2+(\sqrt{\Delta_f^2+g^2M_{mn}^2}+\mu/2)^2\}
		\right]
							\nonumber\\
	&-\tr\ln\left[
		1+
	\frac{4\mu^2g^2M_{mn}^2(\omega_0^2+\Delta_s^2-\mu^2+g^2M_{mn}^2)}
	{(\omega_0^2+\Delta_{A_0}^2+g^2M_{mn}^2)
	\{\omega_0^2+(\sqrt{\Delta_s^2+g^2M_{mn}^2}-\mu)^2\}
	\{\omega_0^2+(\sqrt{\Delta_s^2+g^2M_{mn}^2}+\mu)^2\}}
		\right]
						\nonumber\;.
\end{align}
Let us define all the quantities appearing in the above expression.
The traces are taken over the momentum and the color space. The primed
trace operator implies that the zero spatial momentum is excluded
in the trace. This is because the operators are resulting from the
functional determinants of $x,y$ fields and we are specifically excluding
the zero modes of those fields from the integration. The zeroth
component of the momentum is denoted by $\omega_0$ and
$\Delta^2$'s denote the eigenvalues of the Laplacians for the
$S^3$ spherical harmonics. The specific values of $\Delta^2$'s 
are shown in the
table below.
\\
\\
\centerline{
\begin{tabular}{c|c c c}
Fields    & Symbol           & Eigenvalue & Degeneracy \\ \hline
$A_\iota$ & $\Delta_g^2$     & $(h+1)^2$  & $h(h+2)$	\\ 
$\Phi_A$  & $\Delta_s^2$     & $(h+1)^2$  & $(h+1)^2$	\\
$A_0$     & $\Delta_{A_0}^2$ & $h(h+2)$   & $(h+1)^2$	\\
$x,y$     & $\Delta_s^2$     & $(h+1)^2$  & $(h+1)^2$	\\
$\psi_i$  & $\Delta_f^2$     & $(h+1/2)^2$& $h(h+1)$
\end{tabular}}
\\
\\
The $h$ are non-negative integers.
The eigenvalue $\Delta_s^2$ includes the mass factor, \textit{i.e.},
$\Delta_s^2=h(h+2)+1=(h+1)^2$. Since the $A_0$-scalar does not
have mass, the eigenvalue is $h(h+2)$. At each value of $h$, 
the spherical harmonics are degenerate and the degeneracy factors
are shown in the table and they are assumed in
the trace operators.

Before we proceed to compute the momentum integrals, we comment
on the ghost contributions.
The spatial components of the
gauge fields can be decomposed into the gradient of a scalar
and a divergenceless vector.
The ghosts, which are the scalars,
cancel the former piece in the decomposition and the scalar
$A_0$ contributions.

Now we are going to integrate the momenta. Consider the
generic expression,
\begin{equation*}
	I=\int_{-\infty}^{\infty}\frac{d\omega_0}{2\pi}
	\ln(\omega_0^2+M^2)
					\;,
\end{equation*}
where we assume $M>0$. Then,
\begin{equation*}
	\frac{\partial I}{\partial M^2}
	=\int_{-\infty}^{\infty}\frac{d\omega_0}{2\pi}
	\frac{1}{\omega_0^2+M^2}
	=\frac{1}{2\sqrt{M^2}}
					\;.
\end{equation*}
Integrating back $M^2$, we obtain $I=M$. This result tells us
that the chemical potentials appearing in the third and fourth
lines of (\ref{TraceLog}) cancel and this is interpreted as the
cancellation between the particle and anti-particle pairs.
Thus, except for the last line of (\ref{TraceLog}), the
computations are similar and we demonstrate the
explicit computation for the first term in (\ref{TraceLog}).

We have,
\begin{align*}
	&-\tr\ln(\omega_0^2+\Delta_g^2+g^2M_{mn}^2)
	=-\tr\sqrt{\Delta_g^2+g^2M_{mn}^2}
						\nonumber\\
	&=-\tr\left(\Delta_g+\frac{g^2M_{mn}^2}{2\Delta_g}
		+\cdots	\right)
						\;.
\end{align*}
In the first equality, the $\omega_0$ integration has been done
and in the second equality, the square root has been expanded
with respect to $g^2M_{mn}^2$.
The first term in the last line gives the Casimir energy of the
$S^3$ and we are going to neglect this. We also concentrate on
the term of $g^2M_{mn}^2$ and drop the higher order terms. The
second term in the last line is divergent and we adopt
the dimensional regularization which boils down to the usual
$\zeta$-function regularization. However, as we will see, the
dimensional regularization has the advantage that it nicely
picks up the divergent $1/\epsilon$ pole. The idea is to
extend the spatial $S^3$ manifold to $S^3\times\mathbb{R}^\epsilon$
and use the following replacements,
\begin{align*}
	\sum_h&\rightarrow
	\sum_h\left(\frac{e^{\gamma_E}\Lambda^2}{4\pi}\right)^\epsilon
	\int\frac{d^{-2\epsilon}p}{(2\pi)^\epsilon}
							\;,\nonumber\\
	\Delta_g&\rightarrow \sqrt{p^2+(h+1)^2}
							\;,
\end{align*}
where we adopted the $\overline{MS}$-scheme and $\gamma_E$ is the
Euler's constant and $\Lambda$ is an arbitrary renormalization scale
measured in the units of $1/R$.

The regularization gives,
\begin{align*}
	\tr\frac{1}{\Delta_g}
	&=\frac{1}{2\pi^2}
	\sum_{h=0}^{\infty}h(h+2)\left(\frac{e^{\gamma_E}\Lambda^2}{4\pi}\right)^\epsilon
	\int\frac{d^{-2\epsilon}p}{(2\pi)^\epsilon}
	\frac{1}{\sqrt{p^2+(h+1)^2}}
								\nonumber\\
	&=\frac{1}{2\pi^2}
	\sum_{h=0}^{\infty}h(h+2)\left(e^{\gamma_E}\Lambda^2\right)^\epsilon
	\frac{\Gamma(\epsilon+1/2)}{\Gamma(1/2)}
	\frac{1}{(h+1)^{1+2\epsilon}}
								\nonumber\\
	&=\frac{1}{2\pi^2}
	\left(e^{\gamma_E}\Lambda^2\right)^\epsilon
	\frac{\Gamma(\epsilon+1/2)}{\Gamma(1/2)}
	\left\{\zeta(-1+2\epsilon,1)-\zeta(1+2\epsilon,1)\right\}
								\nonumber\\
	&=\frac{1}{2\pi^2}\left\{
	-\frac{1}{2\epsilon}-\frac{1}{12}-\gamma_E
	+\ln2-\frac{1}{2}\ln\Lambda^2
	+\mathcal{O}(\epsilon)
	\right\}
							\;,
\end{align*}
where we have the generalized $\zeta$-function defined by,
\begin{equation*}
	\zeta(x,y):=\sum_{h=0}^\infty\frac{1}{(h+y)^x}
							\;,
\end{equation*}
with $y\neq0$. Other terms in (\ref{TraceLog}) except the last line
can be evaluated similarly and they give the contribution,
\begin{equation}\label{NiceTerm}
	-\frac{1}{\pi^2}g^2\left(\sum_{m<n}M_{mn}^2\right)
	\left(-\frac{3}{4}+\ln2\right)
					\;.
\end{equation}
We remind the reader that one of the spatial gauge field contributions
and $A_0$ contribution have been canceled by the ghosts and the
primed trace appearing in (\ref{TraceLog}) is evaluated from
$h=1$, excluding the light mode of the $x,y$ fields.
In the expression (\ref{NiceTerm}), the $1/\epsilon$ poles
and the renormalization scheme factors (such as $\gamma_E$ and
$\Lambda$) canceled among the bosonic and fermionic contributions.
\\

\begin{figure}
\centerline{\scalebox{.5}{\includegraphics{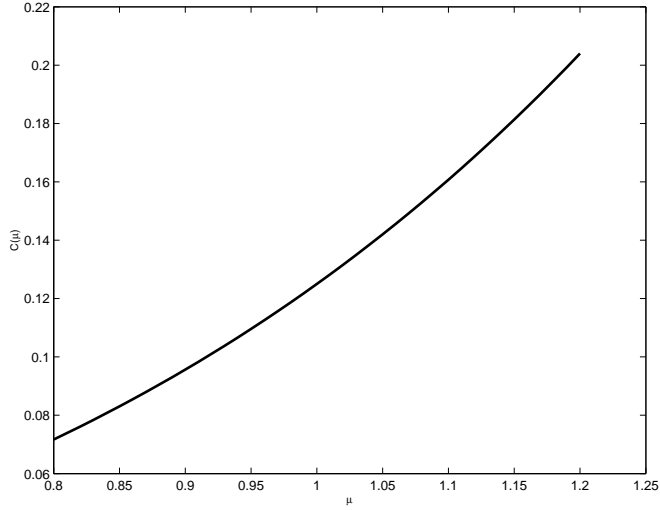}}}
\caption{\footnotesize The plot of the function $C(\mu)$
near the value $\mu=1$.}
\label{CmuGraph}
\end{figure}

We are now going to evaluate the last line in (\ref{TraceLog}).
We expand the logarithm with respect to $g^2M_{mn}^2$ and calculate the first
term in the expansion.
The $\omega_0$ integral can be done and yields,
\begin{equation*}
	-\tr4\mu^2g^2M_{mn}^2
	\left(
	\frac{\Delta_{A_0}+\Delta_s}
	{2\Delta_{A_0}\Delta_s(\Delta_{A_0}+\Delta_s-\mu)
				(\Delta_{A_0}+\Delta_s+\mu)}
	\right)
						\;.
\end{equation*}
The argument of the parenthesis
is logarithmically divergent and asymptotically approaches
to $1/4h$ as $h\rightarrow\infty$. What we are going to do is to
subtract $1/4\Delta_s$ from the above expression and add the
same factor $1/4\Delta_s$. The above expression subtracted by
$1/4\Delta_s$ gives a finite sum and we dimensionally
regulate the added $1/4\Delta_s$.
So,
\begin{align}\label{ComplicatedTerm}
	&-4\mu^2g^2\left(\sum_{m<n}M_{mn}^2\right)\frac{1}{2\pi^2}\sum_{h=1}^{\infty}
	\bigg[
	\left\{
	(h+1)^2\frac{\Delta_{A_0}+\Delta_s}
	{2\Delta_{A_0}\Delta_s(\Delta_{A_0}+\Delta_s-\mu)
				(\Delta_{A_0}+\Delta_s+\mu)}
				-\frac{1}{4\Delta_s}
	\right\}
							\nonumber\\
	&\qquad\qquad\qquad\qquad\qquad\qquad\qquad
	+\left(\frac{e^{\gamma_E}\Lambda^2}{4\pi}\right)^\epsilon
	\int\frac{d^{-2\epsilon}p}{(2\pi)^\epsilon}
	\frac{1}{4\sqrt{p^2+(h+1)^2}}
	\bigg]
							\nonumber\\
	&=-\frac{1}{\pi^2}g^2\left(\sum_{m<n}M_{mn}^2\right)
	\left\{
	C(\mu)
	+\frac{\mu^2}{4\epsilon}
	+\frac{\mu^2}{4}(-2+2\gamma_E-2\ln2+\ln\Lambda^2)
	\right\}
							\;,
\end{align}
where we have defined a monotonically increasing function $C(\mu)$
as,
\begin{equation*}
	C(\mu):=2\mu^2\sum_{h=1}^\infty
		\left\{
		(h+1)^2\frac{\Delta_{A_0}+\Delta_s}
		{2\Delta_{A_0}\Delta_s(\Delta_{A_0}+\Delta_s-\mu)
		(\Delta_{A_0}+\Delta_s+\mu)}
		-\frac{1}{4\Delta_s}
		\right\}
						\;.
\end{equation*}
The behavior of this function around the value $\mu=1$ is plotted
in Figure \ref{CmuGraph}.

The $1/\epsilon$ pole present in (\ref{ComplicatedTerm}) is subtracted
by the mass renormalization counterterm associated with the
chemical potential. To see this, consider the diagram below:
\\
\centerline{\includegraphics{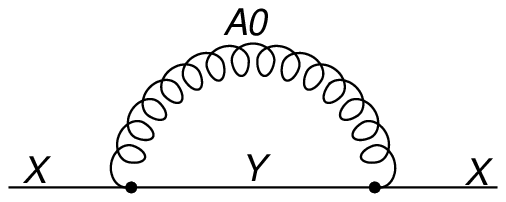}}
\\
The vertices arise from the factor in
the Lagrangian (\ref{FullLagrangian}),
\begin{equation*}
	2\mu g\tr(A_0[X,Y])=2i\mu gf^{abc}A_0^aX^bY^c
							\;.
\end{equation*}
The diagram is logarithmically divergent and in the zero external
momentum limit, it yields the $1/\epsilon$ pole,
\begin{equation*}
	-\frac{\lambda\mu^2}{4\pi^2\epsilon}
							\;,
\end{equation*}
where $\lambda$ is the 't Hooft coupling $\lambda:=C_Ag^2$ with
the second Casimir $C_A$ defined as $f^{abc}f^{dbc}=C_A\delta^{ad}$.
Now observe that,
\begin{equation*}
	\sum_{m<n}\mu^2g^2M_{mn}^2
	=\frac{1}{2}\lambda\mu^2\sum_{a=1}^{N^2-1}\left\{(X^a)^2+(Y^a)^2\right\}
							\;,
\end{equation*}
where we set the generator of the $U(1)$ part of the $U(N)$ gauge
group, $T^{N^2}=\mathbf{1}_{N\times N}/\sqrt{N}$, to the last $N^2$th
one.
We are considering the 
fields that take values in the flat directions, so we have 
$X^a,Y^a\neq0$ only in the Cartan subalgebra valued fields and
the $U(1)$ part is excluded
since the corresponding $U(1)$ field is free and does not contribute
to the one-loop correction.
Thus,
we see that the subtraction of the $1/\epsilon$ pole in fact
comes from the mass renormalization counterterm associated 
with the chemical potential.
We note that this is not the field strength renormalization of
$X,Y$ fields because the divergent diagram considered
above is not proportional to the external momentum and it is
proportional to $\mu^2$.

The one-loop contribution to the effective potential is the
negative of (\ref{NiceTerm}) and (\ref{ComplicatedTerm})
with $1/\epsilon$ pole subtracted. The result is (\ref{OneLoopPtnl}).



\end{document}